\documentclass[10pt,a4paper]{article}
\usepackage{amsmath}
\usepackage{amssymb}
\usepackage{amsfonts}
\usepackage{amstext}
\usepackage{amsbsy}
\usepackage[mathscr]{eucal}
\usepackage{graphicx}
\usepackage{color}
\newcommand{\beq}{\begin{equation}}
\newcommand{\eeq}{\end{equation}}
\newcommand{\beqa}{\begin{eqnarray}}
\newcommand{\eeqa}{\end{eqnarray}}
\newcommand{\ket}[1]{| #1 \rangle}
\newcommand{\bra}[1]{\langle #1 |}

\title{\Large\textbf{Entanglement witnesses and concurrence for  multi-qubit states}}

\author{\textit{ Hoshang Heydari}\\
        \small\textit{Institute of Quantum
Science, Nihon University,}\\
\small\textit{1-8 Kanda-Surugadai, Chiyoda-ku, Tokyo 101-8308,
Japan}
\\\small\textit{Email: hoshang@edu.cst.nihon-u.ac.jp}}
\begin{document}

\maketitle

\begin{abstract}
We establish a relation between concurrence  and entanglement
witnesses. In particular, we construct entanglement witnesses for
three-qubit W and GHZ states in terms of concurrence and different
set of operators that generate it.  We also generalize our
construction for multi-qubit states.
\end{abstract}

\section{Introduction}
Entanglement witnesses are a practical way to detect entangled
states \cite{HHH01}. One recent example of detecting four-photon
entangled state by entanglement witnesses has been reported in
\cite{MB04}. Entanglement witnesses  has a rich geometrical
structure and their construction is a consequence of the Hahn-Banach
theorem.

 On other hand, concurrence is one of well-known  measure of
entanglement. One can also find  different construction and
definition of concurrence for both pure and mixed bipartite and
multipartite states \cite{Albeverio,Mintert}.
 Recently, we have also constructed
 generalized concurrence for pure
general multipartite states based on the complement of a positive
operator valued measure (POVM) on quantum phase \cite{Hosh7}. In
particular, by rewriting orthogonal complement of a POVM on quantum
phase as sums and taking the expectation value of each of these
operators, we were able to construct a general formula for
concurrence. In this paper, we will establish a connection between
entanglement witnesses and concurrence. In section \ref{EW} we will
give short introduction to construction of entanglement witnesses
with some example for three-qubit W and GHZ state. In section
\ref{POVM} we will introduce our construction of concurrence based
on the complement of POVM on quantum phase. And finally, in section
\ref{EWC} we  will show how we can construct entanglement witnesses
for three-qubit states in terms of concurrence and it's generating
operators. Our construction suggests a systematic way of
constructing entanglement witnesses for  multi-qubit states.

 We will consider  a general multipartite quantum
system with $m$ subsystems which we denote as
$\mathcal{Q}=\mathcal{Q}_{1}\mathcal{Q}_{2}\cdots\mathcal{Q}_{m}$,
and denoting its general state as $
\ket{\Psi}=\sum^{N_{1}}_{l_{1}=1}\cdots\sum^{N_{m}}_{l_{m}=1}
\alpha_{l_{1},l_{2},\ldots,l_{m}} \ket{l_{1},l_{2},\ldots,l_{m}}\in
\mathcal{H}_{\mathcal{Q}}=\mathcal{H}_{\mathcal{Q}_{1}}\otimes
\mathcal{H}_{\mathcal{Q}_{2}}\otimes\cdots\otimes\mathcal{H}_{\mathcal{Q}_{m}}$,
where the dimension of the $j$th Hilbert space is given  by
$N_{j}=\dim(\mathcal{H}_{\mathcal{Q}_{j}})$.  Moreover, let
$\rho_{\mathcal{Q}}=\sum^{\mathrm{N}}_{i=1}p_{i}\ket{\Psi_{i}}\bra{\Psi_{i}}$,
for all $0\leq p_{i}\leq 1$ and $\sum^{\mathrm{N}}_{i=1}p_{i}=1$,
denote a density operator acting on the Hilbert space $
\mathcal{H}_{\mathcal{Q}}$.  The most well know examples of
multi-qubit states are $\ket{\Psi_{GHZ^{m}}}$ and
$\ket{\Psi_{W^{m}}}$ state. These quantum states are defined by  $
 \ket{\Psi_{GHZ^{m}}}=\frac{1}{\sqrt{2}}(\ket{1,\ldots,1}+\ket{2,\ldots,2})
$ and
$
\ket{\Psi_{W^{m}}}=\frac{1}{\sqrt{m}}(\ket{m-1,2}),
$
 where $\ket{m-1,2}$ denotes the totally symmetric state including
 $m-1$ ones and 1 twos.
 In the following section, we will call our local operators based
on these classes of states.

\section{Entanglement witnesses}\label{EW}
In this section we will give a short introduction to entanglement
witnesses for general multipartite state. Let $\rho_{\mathcal{Q}}$
be a density operator acting on $\mathcal{H}_{\mathcal{Q}}$. Then,
the density operator $\rho_{\mathcal{Q}}$ is said to be fully
separable, which we will denote by $\rho^{sep}_{\mathcal{Q}}$, with
respect to the Hilbert space decomposition, if it can  be written as
$ \rho^{sep}_{\mathcal{Q}}=\sum^\mathrm{N}_{k=1}p_k
\bigotimes^m_{j=1}\rho^k_{\mathcal{Q}_{j}},~\sum^\mathrm{N}_{k=1}p_{k}=1
$
 for some positive integer $\mathrm{N}$, where $p_{k}$ are positive real
numbers and $\rho^k_{\mathcal{Q}_{j}}$ denotes a density operator on
Hilbert space $\mathcal{H}_{\mathcal{Q}_{j}}$. If
$\rho^{p}_{\mathcal{Q}}$ represents a pure state, then the quantum
system is fully separable if $\rho^{p}_{\mathcal{Q}}$ can be written
as
$\rho^{sep}_{\mathcal{Q}}=\bigotimes^m_{j=1}\rho_{\mathcal{Q}_{j}}$,
where $\rho_{\mathcal{Q}_{j}}$ is the density operator on
$\mathcal{H}_{\mathcal{Q}_{j}}$. If a state is not separable, then
it is said to be an entangled state.

Now, for every entangled state $\rho_{\mathcal{Q}}$ there exist a
Hermitian operator $\mathcal{W}$ that satisfies the following
conditions: i) $\mathrm{Tr}(\mathcal{W}\rho_{\mathcal{Q}})<0$, and
   ii) $\mathrm{Tr}(\mathcal{W}\rho^{sep}_{\mathcal{Q}})\geq0$, for
  all separable state $\rho^{sep}_{\mathcal{Q}}$.
The $\mathcal{W}$ operator usually called entanglement witnesses. By
construction this operator have a positive expectation value on the
set of all separable states. Thus, if the measurement of
$\mathcal{W}$ on a quantum system represented by
$\rho_{\mathcal{Q}}$  produce negative value, then this quantum
$\rho_{\mathcal{Q}}$ is an entangled state. Moreover, the existence
of this operator is guaranteed by the Hahn-Banach theorem: For a
compact and convex set $\mathbb{S}$, if $\rho_{\mathcal{Q}}$ is not
belong to $\mathbb{S}$, then there exists a hyper-plane that
separate $\rho_{\mathcal{Q}}$ from $\mathbb{S}$. Next, we will
construct two well known entanglement witnesses for GHZ and W class
states. For three-qubit GHZ and W states entanglement witnesses can
be constructed as
\begin{equation}
\mathcal{W}_{GHZ^{3}}=\frac{3}{4}\mathbb{I}-\ket{\Psi_{GHZ^{3}}}\bra{\Psi_{GHZ^{3}}},
~\mathcal{W}_{W^{3}}=\frac{2}{3}\mathbb{I}-\ket{\Psi_{W^{3}}}\bra{\Psi_{W^{3}}},
\end{equation}
where $\mathbb{I}$ is an identity matrix. In following section we
will rewrite these entanglement witnesses in terms of concurrence.

\section{Different classes of POVM for general multipartite
states}\label{POVM} In this section, we will construct concurrence
for general pure multipartite states
$\mathcal{Q}^{p}(N_{1},\ldots,N_{m})$, where superscript $p$
indicates that we are only considering pure multipartite states. In
our construction, we will use linear operators that are constructed
by the orthogonal complement of POVM on quantum phase \cite{Hosh7}.
The POVM for each subsystem $\mathcal{Q}_{j}$ is defined by
\begin{eqnarray}
&&\Delta_{\mathcal{Q}_{j}}(\varphi_{k_{j},l_{j}})=
\sum^{N_{j}}_{l_{j},k_{j}=1}
e^{i\varphi_{k_{j},l_{j}}}\ket{k_{j}}\bra{l_{j}},
\end{eqnarray}
where $ \varphi_{k_{j},l_{j}}=
-\varphi_{l_{j},k_{j}}(1-\delta_{k_{j} l_{j}})$. Moreover, the
orthogonal complement of our POVM
 is given by
$\widetilde{\Delta}_{\mathcal{Q}_{j}}(\varphi_{k_{j},l_{j}})=\mathcal{I}_{N_{j}}-
\Delta_{\mathcal{Q}_{j}}(\varphi_{k_{j},l_{j}})$, where
$\mathcal{I}_{N_{j}}$ is the $N_{j}$-by-$N_{j}$ identity matrix for
subsystem $j$.   For $m$-partite quantum system we construct a
operator (matrix) by taking the tensor product of $m$ subsystems as
follows
\begin{eqnarray}
\widetilde{\Delta}_\mathcal{Q}(\varphi_{k_{1},l_{1}},\ldots,
\varphi_{k_{m},l_{m}})&=&
\widetilde{\Delta}_{\mathcal{Q}_{1}}(\varphi_{k_{1},l_{1}})
\otimes\cdots
\otimes\widetilde{\Delta}_{\mathcal{Q}_{m}}(\varphi_{k_{m},l_{m}}),
\end{eqnarray}
where
 $\widetilde{\Delta}_\mathcal{Q}(\varphi_{k_{1},l_{1}},\ldots,
\varphi_{k_{m},l_{m}})$ has phases that are sums or differences of
phases originating from two and $m$ subsystems. That is, in the
latter case the phases of
$\widetilde{\Delta}_\mathcal{Q}(\varphi_{k_{1},l_{1}},\ldots,
\varphi_{k_{m},l_{m}})$ take the form
$(\varphi_{k_{1},l_{1}}\pm\varphi_{k_{2},l_{2}}
\pm\ldots\pm\varphi_{k_{m},l_{m}})$ and identification of these
joint phases makes our distinguishing possible. Thus, we can define
linear operators for the $\mathrm{W}^{m}$ class which are sums and
differences of phases of two subsystems, i.e.,
$(\varphi_{k_{r_{1}},l_{r_{1}}} \pm\varphi_{k_{r_{2}},l_{r_{2}}})$.
That is, for the $\mathrm{W}^{m}$ class we have
\begin{eqnarray}\nonumber
 \widetilde{\Delta}^{
\mathrm{W}^{m}}_{\mathcal{Q}_{r_{1}r_{2}}}
(\Lambda_{m})&=&\mathcal{I}_{N_{1}} \otimes\cdots
\otimes\widetilde{\Delta}_{\mathcal{Q}_{r_{1}}}
(\varphi^{\frac{\pi}{2}}_{k_{r_{1}},l_{r_{1}}}) \otimes\cdots\otimes
\widetilde{\Delta}_{\mathcal{Q}_{r_{2}}}
(\varphi^{\frac{\pi}{2}}_{k_{r_{2}},l_{r_{2}}})\otimes\cdots\otimes
\mathcal{I}_{N_{m}},\\
\end{eqnarray}
where $1\leq r_{1}<r_{2}\leq m$ and the notation
$\widetilde{\Delta}_{\mathcal{Q}_{j}}
(\varphi^{\frac{\pi}{2}}_{k_{j},l_{j}})$ means that we evaluate
$\widetilde{\Delta}_{\mathcal{Q}_{j}}(\varphi_{k_{j},l_{j}})$ at
$\varphi_{k_{j},l_{j}}=\pi/2$ for all $k_{j},l_{j}$. In order to
simplify our presentation, we have used
$(\Lambda_{m})=(k_{1},l_{1};$ $\ldots;k_{m},l_{m})$ as an abstract
multi-index notation.
Next, we could write the linear operator
$\widetilde{\Delta}^{\mathrm{W}^{m}}_{\mathcal{Q}_{r_{1}r_{2}}}(\Lambda_{m})$
as a direct sum of the upper and lower anti-diagonal
\begin{eqnarray}
 \widetilde{\Delta}^{
\mathrm{W}^{m} }_{\mathcal{Q}_{r_{1}r_{2}}}(\Lambda_{m})
&=&\mathfrak{U}\widetilde{\Delta}^{
\mathrm{W}^{m}}_{\mathcal{Q}_{r_{1}r_{2}}}(\Lambda_{m})+\mathfrak{L}\widetilde{\Delta}^{
\mathrm{W}^{m} }_{\mathcal{Q}_{r_{1}r_{2}}}(\Lambda_{m}).
\end{eqnarray}
 For the $\mathrm{GHZ}^{m}$ class, we define linear
operators based on our POVM which are sums and differences of phases
of $m$-subsystems, i.e., $(\varphi_{k_{r_{1}},l_{r_{1}}}
\pm\varphi_{k_{r_{2}},l_{r_{2}}}\pm
\ldots\pm\varphi_{k_{m},l_{m}})$. That is, for the
$\mathrm{GHZ}^{m}$ class we have
\begin{eqnarray}
 \widetilde{\Delta}^{
\mathrm{GHZ}^{m}}_{\mathcal{Q}_{r_{1}r_{2}}}(\Lambda_{m})
&=&\widetilde{\Delta}_{\mathcal{Q}_{1}}
(\varphi^{\pi}_{k_{1},l_{1}})\otimes\cdots
\otimes\widetilde{\Delta}_{\mathcal{Q}_{r_{1}}}
(\varphi^{\frac{\pi}{2}}_{k_{r_{1}},l_{r_{1}}})\\\nonumber&&
\otimes\cdots\otimes \widetilde{\Delta}_{\mathcal{Q}_{r_{2}}}
(\varphi^{\frac{\pi}{2}}_{k_{r_{2}},l_{r_{2}}})\otimes\cdots\otimes
\widetilde{\Delta}_{\mathcal{Q}_{m}} (\varphi^{\pi}_{k_{m},l_{m}}),
\end{eqnarray}
where $\widetilde{\Delta}_{\mathcal{Q}_{j}}
(\varphi^{\pi}_{k_{j},l_{j}})$ indicates that we evaluate
$\widetilde{\Delta}_{\mathcal{Q}_{j}}(\varphi_{k_{j},l_{j}})$ at
$\varphi_{k_{j},l_{j}}=\pi$ for all $k_{j},l_{j}$. Note also that,
in this case we get an operator which has the structure of the Pauli
operator $\sigma_{x}$ embedded in a higher-dimensional Hilbert space
and coincides with $\sigma_{x}$ for a single-qubit. There are
$\frac{m(m-1)}{2} $ linear operators for the $\mathrm{GHZ}^{m}$
class. In our recent paper \cite{Hosh7} we have construct
concurrence for general multipartite states based on these sets of
operators which is given by
\begin{eqnarray}\nonumber
\mathcal{C}(\ket{\Psi})&=&
    (\mathcal{N}_{m}\{\sum_{1\leq r_{1}<r_{2}\leq m}\mathcal{C}(\mathcal{Q}^{W^{m}}_{r_{1}r_{2}}
    )
    +\sum_{1\leq r_{1}<r_{2}\leq m}\mathcal{C}(\mathcal{Q}^{GHZ^{m}}_{r_{1}r_{2}}
    )+\ldots\})^{1/2},
\end{eqnarray}
where $\mathcal{N}_{m}$ is a normalization constant. The definition
of these terms can be fund in the above mentioned paper. For
three-qubit state it is given by
\begin{eqnarray}\nonumber
\mathcal{C}(\ket{\Psi})&=&( 2|
\alpha_{1,1,1}\alpha_{2,2,1}-\alpha_{1,2,1}\alpha_{2,1,1} |^{2}
+2|\alpha_{1,1,2}\alpha_{2,2,2}-\alpha_{1,2,2}\alpha_{2,1,2}|^{2}\\\nonumber&&+
2| \alpha_{1,1,1}\alpha_{2,1,2}-\alpha_{1,1,2}\alpha_{2,1,1}|^{2}
 +2|\alpha_{1,2,1}\alpha_{2,2,2}-\alpha_{1,2,2}\alpha_{2,2,1}|^{2}\\\nonumber&&+
 2|\alpha_{1,1,1}\alpha_{1,2,2}-\alpha_{1,1,2}\alpha_{1,2,1}|^{2}+
2|\alpha_{2,1,1}\alpha_{2,2,2}-\alpha_{2,1,2}\alpha_{2,2,1}|^{2}\\\nonumber&&
+| \alpha_{1,1,1}\alpha_{2,2,2}-\alpha_{1,1,2}\alpha_{2,2,1}|^{2}+
|\alpha_{1,1,1}\alpha_{2,2,2}- \alpha_{1,2,1}\alpha_{2,1,2} |^{2}
\\\nonumber&&+
| \alpha_{1,1,1}\alpha_{2,2,2}- \alpha_{1,2,2}\alpha_{2,1,1} |^{2} +
|\alpha_{1,1,2}\alpha_{2,2,1}-
\alpha_{1,2,1}\alpha_{2,1,2}|^{2}\\\nonumber&&
+ | \alpha_{1,1,2}\alpha_{2,2,1}- \alpha_{1,2,2}\alpha_{2,1,1}
|^{2}+ |\alpha_{1,2,1}\alpha_{2,1,2}- \alpha_{1,2,2}\alpha_{2,1,1}
 |^{2})^{\frac{1}{2}},
\end{eqnarray}
where we have set $\mathcal{N}=1/4$. Next, we evaluate this measure
for $\ket{\Psi_{W^{3}}}$ and $\ket{\Psi_{GHZ^{3}}}$ states. For
these two well-known states we have
$\mathcal{C}^{2}(\ket{\Psi_{W^{3}}})=2\cdot3\frac{1}{9}=\frac{2}{3}$
and $\mathcal{C}^{2}(\ket{\Psi_{GHZ^{3}}})=\frac{3}{4}$
respectively.
\section{Entanglement witnesses based on concurrence}\label{EWC}
Now we will systematically construct entanglement witnesses for
multipartite states. But before that we need to introduce some new
notations.  For $m$-partite quantum system we will denote by $
\widetilde{\Delta}^{\pm}_\mathcal{Q}(\varphi_{k_{1},l_{1}},\ldots,
\varphi_{k_{m},l_{m}}) $ the operator which has only phases that
either can be written as $(+\varphi_{k_{1},l_{1}}
+\varphi_{k_{2},l_{2}}+ \ldots +\varphi_{k_{m},l_{m}})$ or
 $(-\varphi_{k_{1},l_{1}} -\varphi_{k_{2},l_{2}}-
\ldots -\varphi_{k_{m},l_{m}})$. This means the sign in front of all
these phases is either positive or negative. The elements of $
\widetilde{\Delta}^{+}_\mathcal{Q}(\varphi_{k_{1},l_{1}},\ldots,
\varphi_{k_{m},l_{m}}) $ and $
\widetilde{\Delta}^{-}_\mathcal{Q}(\varphi_{k_{1},l_{1}},\ldots,
\varphi_{k_{m},l_{m}}) $ are placed over and under main diagonal
respectively. So we can also rewrite our operators as
\begin{eqnarray}
\widetilde{\Delta}^{\pm}_\mathcal{Q}(\varphi_{k_{1},l_{1}},\ldots,
\varphi_{k_{m},l_{m}})=\widetilde{\Delta}^{+}_\mathcal{Q}(\varphi_{k_{1},l_{1}},\ldots,
\varphi_{k_{m},l_{m}})+\widetilde{\Delta}^{-}_\mathcal{Q}(\varphi_{k_{1},l_{1}},\ldots,
\varphi_{k_{m},l_{m}}).
\end{eqnarray}
For example for three-qubit states we have two classes of these
operators, that is W class and GHZ class operators. For W class we
have
\begin{eqnarray}\label{W3}\nonumber
 \widetilde{\Delta}^{
W^{3}\pm}_{\mathcal{Q}_{1,2}}(\Lambda_{3})
&=&\left(\widetilde{\Delta}_{\mathcal{Q}_{1}}
(\varphi^{\frac{\pi}{2}}_{1,2})
\otimes\widetilde{\Delta}_{\mathcal{Q}_{2}}
(\varphi^{\frac{\pi}{2}}_{1,2})\otimes \mathcal{I}_{2}\right)^{\pm}.
\end{eqnarray}
$
 \widetilde{\Delta}^{
W^{3}\pm}_{\mathcal{Q}_{1,3}} (\Lambda_{3})$ and $
 \widetilde{\Delta}^{
W^{3}\pm}_{\mathcal{Q}_{2,3}}(\Lambda_{3}) $ are defined
 in the similar way. Moreover, for three-qubit GHZ we have
 \begin{eqnarray}\nonumber
\widetilde{\Delta}^{ GHZ^{3}\pm}_{\mathcal{Q}_{1,2}}&=&\left(
\widetilde{\Delta}_{\mathcal{Q}_{1}} (\varphi^{\frac{\pi}{2}}_{1,2})
\otimes\widetilde{\Delta}_{\mathcal{Q}_{2}}
(\varphi^{\frac{\pi}{2}}_{1,2})\otimes
\widetilde{\Delta}_{\mathcal{Q}_{3}}
(\varphi^{\pi}_{1,2})\right)^{\pm}.
\end{eqnarray}
 Now, we are in a position to rewrite entanglement
witnesses for three-qubit states in terms of concurrence and it's
operators. For example for a three-qubit GHZ  state the entanglement
witnesses can be constructed as
\begin{equation}
\mathcal{W}_{GHZ^{3}}=\frac{3}{4}\mathbb{I}-\ket{\Psi_{GHZ^{3}}}\bra{\Psi_{GHZ^{3}}}=
\mathcal{C}^{2}(\ket{\Psi_{GHZ^{3}}})\mathbb{I}-\ket{\Psi_{GHZ^{3}}}\bra{\Psi_{GHZ^{3}}}.
\end{equation}
We can also go one step further by setting
$\mathcal{C}^{2}(\ket{\Psi_{GHZ^{3}}})=\mathcal{C}_{g}=\overline{\mathcal{C}}_{g}+1$,
and diagonal matrix
$
   \mathcal{D}_{g}=diag(\overline{\mathcal{C}}_{g},\mathcal{C}_{g},\ldots,\mathcal{C}_{g},\overline{\mathcal{C}}_{g})
$
then an entanglement witnesses for three-qubit GHZ state can be
written as
\begin{eqnarray}
\mathcal{W}_{GHZ^{3}}&=& \mathcal{D}_{g}-\widetilde{\Delta}^{
GHZ^{3}\pm}_{\mathcal{Q}_{1,2}}(\Lambda_{3}).
\end{eqnarray}
For three-qubit  W state entanglement witnesses can be constructed
as
\begin{eqnarray}
\mathcal{W}_{W^{3}}&=&\frac{2}{3}\mathbb{I}-\ket{\Psi_{W^{3}}}\bra{\Psi_{W^{3}}}=
\mathcal{C}^{2}(\ket{\Psi_{W^{3}}})\mathbb{I}-\ket{\Psi_{W^{3}}}\bra{\Psi_{W^{3}}}
\end{eqnarray}
Now, let
$\mathcal{C}^{2}(\ket{\Psi_{W^{3}}})=\mathcal{C}_{w}=\overline{\mathcal{C}}_{w}+1$,
and  define diagonal matrix
$
   \mathcal{D}_{w}=diag(\mathcal{C}_{w},\overline{\mathcal{C}}_{w},$ $\overline{\mathcal{C}}_{w},
   \mathcal{C}_{w},\overline{\mathcal{C}}_{w},\mathcal{C}_{w},\mathcal{C}_{w},\mathcal{C}_{w}).
$
Then an entanglement witnesses for three-qubit W state can be
written as
\begin{eqnarray}
\mathcal{W}_{W^{3}}&=&
\mathcal{D}_{w}-\sum_{r<s}\mathfrak{U}\widetilde{\Delta}^{
W^{3}\pm}_{\mathcal{Q}_{r,s}}(\Lambda_{3}).
\end{eqnarray}
These construction suggest that we can generalize entanglement
witnesses for at least multipartite W and GHZ states. For example
the entanglement witnesses for multi-qubit W states is given by
\begin{eqnarray}
\mathcal{W}_{W^{m}}&=&\gamma\mathbb{I}-\ket{\Psi_{W^{m}}}\bra{\Psi_{W^{m}}}=
\mathcal{C}^{2}(\ket{\Psi_{W^{m}}})\mathbb{I}-\ket{\Psi_{W^{m}}}\bra{\Psi_{W^{m}}}
\\\nonumber&=&\mathcal{D}_{w}-\sum_{r<s}\mathfrak{U}\widetilde{\Delta}^{
W^{m}\pm}_{\mathcal{Q}_{r,s}}(\Lambda_{m}),
\end{eqnarray}
where $\mathcal{D}_{w}$ is a diagonal matrix with elements
$\overline{\mathcal{C}}_{w}=\mathcal{C}^{2}(\ket{\Psi_{W^{m}}})-1$
for non zero coefficients $\alpha_{l_{1},l_{2},\ldots,l_{m}}\neq0$
of multi-qubit W state and
$\mathcal{C}_{w}=\mathcal{C}^{2}(\ket{\Psi_{W^{m}}})$ when
$\alpha_{l_{1},l_{2},\ldots,l_{m}}=0$ for multi-qubit W state. We
have already seen construction of this diagonal matrix for three
qubit states.

Moreover, an entanglement witnesses for multi-qubit GHZ states is
given by
\begin{eqnarray}\nonumber
\mathcal{W}_{GHZ^{m}}&=&\gamma\mathbb{I}-\ket{\Psi_{GHZ^{m}}}\bra{\Psi_{GHZ^{m}}}=
\mathcal{C}^{2}(\ket{\Psi_{GHZ^{m}}})\mathbb{I}-\ket{\Psi_{GHZ^{m}}}\bra{\Psi_{GHZ^{m}}}
\\&=&\mathcal{D}_{g}-\widetilde{\Delta}^{
GHZ^{m}\pm}_{\mathcal{Q}_{1,2}}(\Lambda_{m}),
\end{eqnarray}
where $\mathcal{D}_{g}$ is defined in the same way as for W state,
e.g., by replacing W state by GHZ state. However, we need carefully
chose the normalization constant $\mathcal{N}_{m}$ in the expression
for concurrence. We also can write a general formula for
entanglement witnesses of pure multi-qubit state in terms of
concurrence
\begin{eqnarray}
\mathcal{W}_{\Psi}&=&\gamma\mathbb{I}-\ket{\Psi}\bra{\Psi}=
\mathcal{C}^{2}(\ket{\Psi})\mathbb{I}-\ket{\Psi}\bra{\Psi}
=\mathcal{D}-\sum_{r<s}\widetilde{\Delta}^{
\Psi\pm}_{\mathcal{Q}_{r,s}}(\Lambda_{m}),
\end{eqnarray}
where $\mathcal{D}$ is a diagonal matrix with elements
$\overline{\mathcal{C}}=\mathcal{C}^{2}(\ket{\Psi})-1$ for non zero
coefficients $\alpha_{l_{1},l_{2},\ldots,l_{m}}\neq0$  of the state
$\ket{\Psi}$  and $\mathcal{C}=\mathcal{C}^{2}(\ket{\Psi})$ whenever
$\alpha_{l_{1},l_{2},\ldots,l_{m}}=0$ for given state $\ket{\Psi}$.
For GHZ states we only need to consider $r=1$ and $s=2$ without any
partitions. Thus, it is possible to systematically construct
entanglement witnesses for multi-qubit states in terms of
concurrence and the operators which generate it. The geometrical
structure of entanglement witnesses usually related to geometry of
Hilbert space in functional analysis. On the other hand the geometry
of concurrence and completely separable set is given by the Segre
variety which is defined to be the image of the Segre embedding.
Thus these connection between entanglement witnesses and concurrence
also suggest some, not yet known, similarity between geometrical
structure of these entanglement tools that distinguish between
separable and entangled set of multi-qubit states.
\begin{flushleft}
\emph{Acknowledgments:} The  author acknowledges the financial
support of the Japan Society for the Promotion of Science (JSPS).
\end{flushleft}



\begin{thebibliography}{99}
\bibitem{HHH01} M. Horodecki, P. Horodecki, and R. Horodecki, Phys. Lett. A  283, 1
(2001).
\bibitem{MB04} M. Bourennane, M. Eibl, C. Kurtsiefer, H. Weinfurter, O.
G\"{u}hne, P. Hyllus, D. Bru$\ss$, M. Lewenstein, and A. Sanpera,
Phys. Rev. Lett. {\bf 92} 087902 (2004)
\bibitem{Albeverio} S. Albeverio and S. M. Fei, J. Opt. B: Quantum Semiclass.
Opt. {\bf 3}, 223 (2001).
\bibitem{Mintert}  F. Mintert, M. Kus, and A.Buchleitner, Phys. Rev. Lett. {\bf95}, 260502 (2005).
\bibitem{Hosh7} H. Heydari, J. Phys. A: Math. Gen. {\bf 39} (2006) 15225-15229
\end{thebibliography}
\end{document}